# The Potential of Olfactory Stimuli in Stress Reduction through Virtual Reality

Yasmin Elsaddik Valdivieso, Mohd Faisal, Karim Alghoul, Monireh (Monica) Vahdati,
Kamran Gholizadeh Hamlabadi, Fedwa Laamarti, Hussein Al Osman, Abdulmotaleb El Saddik
*MCRLab, School of Electrical Engineering and Computer Science*
*University of Ottawa,* Ottawa, Canada
(yelsa104, mmohd055, karim.Alghoul, Monica.vhd, kamran.gh, flaamart, hussein.alosman, elsaddik)@uottawa.ca

*Abstract*—Immersive virtual reality (VR) is a promising tool for stress reduction and relaxation, traditionally relying on visual and auditory stimuli. This study examines the role of olfactory stimuli in enhancing these effects, using a randomized within-subject design. Thirty participants aged 18–60 experienced VR scenarios simulating a calming seaside environment, with sessions lasting 45 minutes, in two conditions: with and without a "Beach" essential oil scent (Yankee Candle) administered via diffuser. Stress and relaxation were assessed through self-reported surveys and physiological measures, specifically ECG-based heart rate variability (HRV). Results showed no significant difference in self-reported relaxation scores ($p=0.371$) between conditions, but HRV analysis revealed a significant stress reduction ($p=0.002$) with olfactory input, with HF increasing 108% from the Math Stress Test to the scented relaxation condition, compared to 44% without scent. Additionally, 71.4% of participants expressed willingness to use olfactory-enhanced VR for relaxation, suggesting practical appeal. These findings indicate that olfactory stimuli may enhance relaxation subconsciously, underscoring the importance of multisensory integration in VR. Future work could explore personalized scents and long-term effects to optimize VR-based interventions for emotional and physical well-being.

*Index Terms*—Virtual Reality, Olfactory Stimuli, Stress Reduction, Multisensory Integration, Relaxation, Immersive Environments, User Experience

I. INTRODUCTION

Research has shown that being immersed in a natural environment can help reduce stress and anxiety, with studies highlighting both psychological and physiological benefits [1]. Exposure to nature has been linked to changes in physiological signals, such as heart rate and cortisol levels, which are indicative of reduced stress [2]. However, not everyone has access to these natural environments, making it challenging for many to experience these stress-relieving effects in their daily lives.

Immersive VR can be used as an alternative by simulating real-life environments or situations, enabling users to feel present by interacting with the environment. VR technologies provide a sense of "being there" without being physically present, replacing their real surroundings with a virtual version. Users experience immersion in virtual environments primarily through the stimulation of their senses. To enhance sensory experiences and deepen the sense of presence, researchers have introduced various sensory tools into VR, such as touch, smell, and hearing [3]. Affecting users' senses, VR can closely mimic real-world interactions. While VR is commonly associated with gaming, studies have demonstrated its potential to promote relaxation, and reducing anxiety [4].

The ability to smell is mediated by olfactory sensory neurons, which transmit signals to the olfactory bulb—a brain structure with direct connections to the amygdala and hippocampus, regions responsible for processing emotions and memories. Engaging the sense of smell in VR environments may provide a unique pathway to enhance mood and emotional well-being. The concept of using scents to influence emotional well-being is called aromatherapy, which involves using essential oils to promote emotional health and wellness [5]. Aromatherapy with specific scents such as lavender and orange is known to reduce anxiety and improve mood [5]. This paper aims to evaluate the effectiveness of olfactory displays in reducing stress and enhancing relaxation, both measured and perceived. To explore this, participants were exposed to VR environments with and without olfactory stimuli. By randomizing the order of the scented and unscented sessions, we aimed to determine whether the addition of olfactory stimuli could enhance relaxation and reduce stress compared to a purely visual and auditory VR experience.

II. RELATED WORK

VR has gained significant traction across various fields, including healthcare and psychology, due to its ability to simulate real-life scenarios and enhance user engagement. Research has explored the interplay between emotions, colors, sounds, and scents in VR environments, finding that their combination can amplify emotional experiences for users [6]. For instance, [7] utilized a system featuring a 360-degree video of a beach, an olfactory necklace, a VR headset, an EEG headband, and a smartphone app to monitor brain activity and assess the users' mental states. Furthermore, exposure to nature scenes in VR has been shown to be an effective method of reducing stress and improving general well-being [8].

An experiment by [9] explored the use of VR to promote behavioral health, discovering that incorporating olfactory stimuli into the VR environment helped reduce negative affect and state anxiety levels. However, their study had a small, homogenous sample (N=10) and lacked physiological measures, limiting result robustness. Another study by [6] found that aroma enhanced immersion and reduced stress, but their study focused on a single scent (lavender) and short exposure duration, with VR content limited to gaming. While these findings are promising, further research is needed to include diverse participants, integrate physiological and subjective measures, explore varied olfactory stimuli, and assess long-term relaxation effects in VR.

Table 1 summarizes various immersive environments found in literature, particularly focusing on studies that incorporate olfactory stimuli. These studies examine the impact of scent as part of a multisensory experience to enhance relaxation, reduce stress, and improve overall immersion. The studies employing olfactory elements—such as lavender and citrus scents—demonstrate measurable improvements in emotional well-being and user engagement. The use of olfactory stimuli, alongside visual and auditory components, enriches virtual experiences, making it an essential factor in the design of high-impact immersive environments for various applications, including healthcare and astronaut training.

## III. Materials and Methods

To study the potential of olfactory stimuli in VR on stress reduction, we designed a system consisting of a VR environment where an individual can explore a scene with visual and auditory stimuli, as well as a diffuser to produce the olfactory stimuli. The VR environment is meant to be relaxing. The visual stimuli include the scene of a beach with waves as shown in Figure 1. And the auditory stimuli consist of the sounds of waves. The olfactory stimuli that we used was that of a beach scent. Our research questions are as follows:

RQ1: Can olfactory displays reduce stress and induce relaxation in Virtual Environments?

RQ2: How efficient are olfactory displays in reducing stress and enhancing relaxation, both measured and perceived?

To answer our research questions, we designed an experiment where participants are invited to experience the VR environment with and without olfactory stimuli. We used a within-subject design, by randomizing the order of the scented and unscented sessions, and we aimed to determine whether the addition of olfactory stimuli can enhance relaxation and reduce stress compared to a purely visual and auditory VR experience.

### A. Physiological

**ECG Belt:** A Polar H10 ECG belt was used in this experiment. Participants wore the belt around their ribcage to measure cardiac electrical activity using two electrode patches. The Polar H10 belt captures R-R intervals at a sample rate of 1000 Hz and transmits the data in real-time via Bluetooth. The R-R intervals represent the time between two consecutive R-peaks in the ECG waveform, which corresponds to the time between two heartbeats and is commonly used for heart rate variability analysis.

### B. Psychological

**PANAS:** Positive and Negative Affect Schedule (PANAS) was used to assess changes in participants' mood. This self-report measure includes a series of adjectives describing positive and negative emotions, which participants rated on a scale.

**SAM:** The Self-Assessment Manikin (SAM) is a pictorial assessment tool which measures emotional responses along three dimensions: valence, arousal, and dominance [14]. Valence is represented by figures ranging from happy to unhappy, while arousal is depicted by figures ranging from excited to relaxed. The dominance dimension, which reflects the participant's sense of control over the situation, is represented by changes in the size of the figures (e.g., a larger figure indicates greater perceived control). Participants were asked to select from any of the five figures on each scale. In this study, we focus on valence and arousal while excluding dominance, as these two labels are more effective in capturing the aspects of relaxation and stress reduction.

**RRS:** Participants' self-reported relaxation levels were assessed using the Relaxation Rating Scale (RRS). This scale, adapted from prior research on stress management and relaxation techniques [14], allows participants to rate their level of relaxation on a scale from 0 to 10. A rating of 0 indicates that the participant is not relaxed at all, while a rating of 10 signifies the highest possible level of relaxation. This simple yet effective scale provided a direct measure of how relaxed participants felt after each VR session.

Additionally, participants answered two final scale questions to assess the overall effectiveness and immersion of the VR experience: "Would you use this method as a relaxation technique?" and "Did you feel more immersed with the scent compared to without the scent?" These questions provided insights into the practical applicability of the olfactory-enhanced VR environment and participants' perceived level of immersion.

## IV. Study Design

Participants were randomly grouped into two experimental places. Group one started VR with scent and then VR environment without scent, group 2 started with VR environment without scent and then VR with scent. In all cases, visual and auditory stimuli were present, while olfactory stimuli were introduced only in the scented condition based on the participant's assigned group.

### A. Participants

The study sample consisted of 30 participants (13 female and 17 male), aged between 18 and 40 years (M = 26.73, SD = 4.54). The inclusion criteria required participants to have no history of seizures, severe motion sickness, epilepsy, or other neurological conditions that could be triggered by the VR simulation [15]. Additionally, participants with known allergies to the scents used in the experiment were excluded. Participants were randomly allocated to one of the two experimental groups. Prior to the experiment, participants were instructed to abstain from alcohol for 12 hours and from caffeine, smoking,

TABLE I
NATURE OF SCENES IN IMMERSIVE ENVIRONMENTS IN LITERATURE RESEARCH

| Ref. | Environments | Measurements | Equipment | Constraints | Result |
|---|---|---|---|---|---|
| [10] 2022 | 360-degree panoramic view of a forest scene, multisensory inputs such as temperature, scent (Olfactory), and sound | Physiological measurements: Skin temperature, BVP, and EDA from the wristband. ECG and breathing curves from the chest strap. EEG, EOG, and EMG from the instrumented VR headset. Users' stress, anxiety, arousal | Biopac MP150 system, Multisensory pod showing A-fans, B-heating elements, C-scent diffuser, Oculus Quest VR headset, Zephyr BioHarness3 chest-strap, Empatica E4 wristband | 16 participants for a duration of 75 minutes | Participants who experienced a 15-minute ultra-reality multisensory nature walk had significantly lower levels of stress and higher levels of relaxation compared to those who watched a relaxation Video. |
| [9] 2022 | Nature-inspired forested scent area (Olfactory), small auditorium | Anxiety levels (pre- and post-intervention), subjective measures through PANAS questionnaires | HTC Vive Pro (VR headset), OSG | 10 participants for a duration of 4 minutes, limited difference between live audience and virtual audience | Adding olfactory stimuli to VR environment reduced negative affect and state anxiety levels. |
| [8] 2020 | Immersive Virtual Environments: 360 x 270-degree videos of real parks and scent (Olfactory) | EDA and HR using Empatica E4 wristbands, VAS for stress, anxiety, insecurity, calmness, happiness, and BEI for plant and animal perceptions | Oculus Rift HMD with scents from "Demeter Fragrance Library" | 52 participants for a duration of 15 minutes, some results did not reach statistical significance | Exposure to biodiverse virtual environments could be effective in reducing stress and improving well-being. |
| [11] 2020 | Video display, sound system, diffuser emitting a citrus scent (Olfactory), Natural light, Comfortable seating | Various metrics and devices including EDA, HRV, EEG, and psychological data analyzed with iMotions software. SUS questionnaire | Ambient music, diffuser emitting a citrus scent, visual elements such as murals and artwork | 120 participants for a duration of 67 minutes, time-limited Arousal Increase in Virtual Reality | A multisensory waiting room environment, including calming videos, soothing music, a diffuser emitting a citrus scent, natural light, and comfortable seating, was associated with reduced stress and anxiety in pediatric patients. |
| [12] 2020 | Tilt Brush by Google, a virtual drawing software program that creates immersive 3-D images in VR and scent (Olfactory) | Quantitative scales for anxiety, stress, affect, self-efficacy, creative agency, satisfaction with life. Qualitative thematic analysis | Tilt Brush (VR drawing software), Windows Mixed-Reality headset, remote control devices, Aeroscena diffuser system | 24 participants for a duration of 1 hour, physical discomfort | VR tools with fragrance stimuli reduced negative affect, improved self-efficacy, and revealed common themes related to nature, fantasy, memories, and everyday objects |
| [13] 2019 | Park, Forest, Urban Area and scent (Olfactory) | Skin conductance level, subjective stress sensibility | VR mask, 2D 360° VR photos, earphones, olfactometer and scent (Olfactory) | 154 participants for a duration of 13 minutes | Green environments have greater stress reduction compared to urban environments |
| [7] 2018 | 360° video recording of a beach and scent (Olfactory) | EEG for brain activity, RRS questionnaire | VR Headset, Olfactory Necklace, EEG Headband, smartphone app, Bluetooth headphones | 12 participants for a duration of 10 minutes | Increased perceived relaxation and Relax score from users' EEG |
| [6] 2018 | VR roller coaster environments and scent (Olfactory) | Measuring brain waves with and without aroma to record and analyze EEG on the frontal lobe. | HTC VIVE PRO VR device, two-channel portable biosignal detector | 10 participants. The study did not consider the influence of user interaction due to the limited motion and absence of a controller, requiring further research. | Aroma enhanced the VR experience and reduced discomfort, stress, and dizziness. Significant changes in EEG data and biosignal values were observed. Further research needed for user interaction impact |

EDA = Electrodermal Activity, HR = Heart Rate, VAS = Visual Analogue Scales, BEI = Biodiversity Experience Index, OSG = Olorama scent generator, SCL = Skin conductance level, BVP = Blood Volume Pulse, STAI-S = Spielberger State-Trait Anxiety Inventory, RRS = Relaxation Rating Scale, ECG = Electrocardiography, EEG = Electroencephalography, EOG = Electrooculography, EMG = Electromyography, HRV = Heart Rate Variability.

eating, and exercise for 3 hours. Recruitment was conducted via posters displayed on the university campus Additional participants were recruited through a snowballing technique, where existing participants were encouraged to share the study information with friends and colleagues.

### B. Equipment

**VR Headset:** We used the Meta Quest headset to create an immersive virtual beach environment. Its high-resolution display and motion tracking allowed participants to explore naturally while seated. The wireless design minimized distractions, ensuring an uninterrupted VR experience.

**Diffuser:** To introduce olfactory stimuli into the VR environment, we used a commercially available humidifier, which functions as a diffuser by dispersing essential oils into the air. To simulate the scent of a beach environment, we added a carefully selected essential oil blend designed to evoke fresh oceanic aromas. The diffuser was activated as soon as the participant entered the experiment room to observe the olfactory condition, ensuring a consistent ambient scent throughout the session. This setup allowed for passive scent exposure without requiring direct interaction from participants, which aligns with the immersive nature of the VR experience.

### C. Procedure

The experiment was conducted in two separate office spaces to ensure a consistent and controlled environment: one infused with the beach scent and the other without any scent. The order in which participants experienced the olfactory stimuli was randomized ensuring that the results were not biased by the sequence of conditions. Each experiment session lasted approximately 45 minutes to induce and measure both stress and relaxation in participants, assessed both quantitatively and qualitatively. For the olfactory stimuli portion of the experiment, we used a "Beach" essential oil (Yankee Candle) to evoke a calming, seaside atmosphere, with plain air serving as the control condition.

Participants began by completing a pre-survey questionnaire, including the PANAS, to establish a baseline for their mood and affective state. Following this, each participant was fitted with an ECG belt. To accurately measure their baseline heart rate variability, participants were first seated in an office chair and were asked to read a provided book, The Silk Roads by Peter Frankopan. Reading was chosen over the use of personal devices to control external variables, ensuring that participants were not inadvertently exposed to stressors from their phones. The use of the same book for all participants further ensured consistency. This reading period lasted 6 minutes, allowing time for a stable baseline to be recorded.

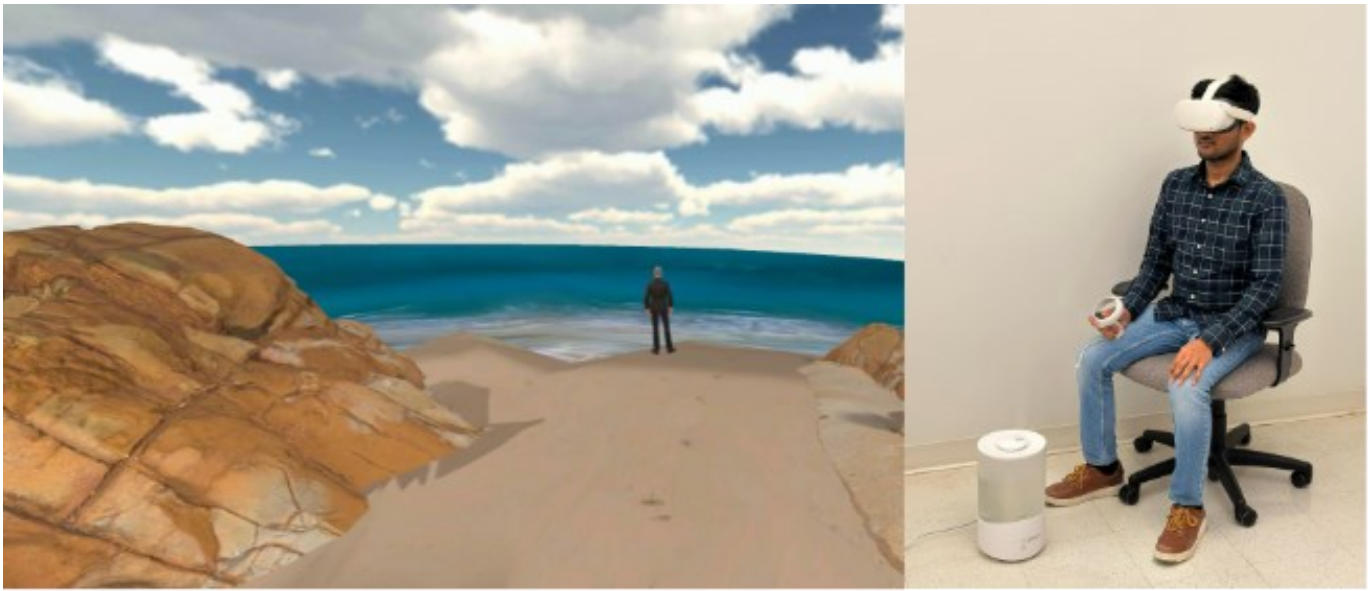

Fig. 1. (Left) Virtual beach environment with an avatar inside the scene. (Right) A participant seated in a chair wearing a Meta Quest VR headset, with a humidifier/diffuser dispersing beach scent to enhance immersion.

Following the baseline measurement, participants completed an arithmetic stress test designed by [16] to induce stress. This task lasted 6 minutes and involved answering consecutive computer-generated multiplication questions involving numbers between 0 and 12 within a set time limit. A score counter was displayed in the upper right corner, and participants' scores decreased with each incorrect answer, creating a pressured environment intended to stimulate stress.

After the arithmetic stress test, participants were fitted with the Meta Quest VR 5 headset and immersed in a virtual beach environment (see Figure 1). For half of the participants, this first VR session included olfactory stimuli, while for the other half, it did not. Participants were encouraged to explore the VR scene at their own pace, turning their heads and interacting with the environment while remaining seated . This immersive experience lasted for 6 minutes, providing ample time to gather ECG data during a relaxed state.

Next, participants completed a second arithmetic stress test identical to the first. This step was necessary to reinduce stress, ensuring that the subsequent VR session could be accurately evaluated for its effectiveness in promoting relaxation.

After the second stress test, participants experienced the opposite VR condition (i.e., those who initially experienced VR with olfactory stimuli now experienced it without, and vice versa). This session also lasted 6 minutes and followed the same procedure as the first VR exposure. By switching the order of VR with and without olfactory stimuli, for different participants, the study ensured that any observed effects were not due to the sequence of conditions but rather the presence or absence of olfactory stimuli in the VR experience.

At the end of each VR session and after the first stress test, participants filled out a post-experiment questionnaire, which included the PANAS, SAM, and RRS scales, to measure changes in mood, stress, and relaxation levels.

## V. RESULTS AND DISCUSSIONS

### A. HRV Analysis

Heart rate variability (HRV) refers to the fluctuations in the time intervals between successive heartbeats. HRV metrics are utilized to evaluate the overall health of the heart and to provide insights into the autonomic nervous system (ANS) status [16]. During relaxation, the parasympathetic nervous system is predominantly activated. Monitoring the balance between the sympathetic and parasympathetic branches allows for inferences regarding the degree of relaxation experienced by an individual.

The assessment of relaxation levels using HRV is primarily conducted in the frequency domain. Of particular interest is the high-frequency (HF) band, which ranges from 0.15 to 0.4 Hz. There is substantial evidence that an increase in the HF component of HRV is associated with increased relaxation levels [17]. The HF component is widely regarded as indicative of parasympathetic (vagal) activity, which plays a key role in promoting relaxation and reducing stress. Furthermore, research has demonstrated that individuals with higher HF baseline tend to exhibit better emotional regulation, lower anxiety, and a better capacity to cope with stress, all of which are indicators of a more relaxed physiological state [17].

For our HRV analysis, we utilized raw ECG signals collected from the Polar H10 belt. The RR intervals of the subjects were accessed and exported using the Android appli-cation "Elite HRV." Afterward, the data was processed using the Python package "hrv-analysis" [18].

Prior to data analysis, RR interval artifacts such as outliers and ectopic beats were identified and removed using the built-in artifact correction methods provided by the library, which apply filtering techniques to clean the signal. Next, we performed frequency domain analysis by calling the "get frequency domain features" function. This function computes HRV parameters using the Welch method for spectral estimation, applying Fast Fourier Transform (FFT) to obtain power across standard frequency bands. In our case, we specifically extracted the HF component, which reflects parasympathetic activity related to relaxation. The high-frequency parameter was then extracted for each of the five parts of each recording session. The average HF values of all the subjects are shown in Figure 2. The percentage increase in HF between the Math stress test and the relaxation experiment was calculated using Equation 1.

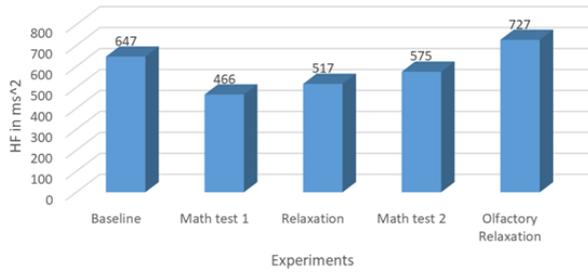

Fig. 2. Mean high-frequency HRV values across all participants

To compare the HF increase under two conditions—relaxation without olfactory stimulation and relaxation with olfactory stimulation—we calculated the average HF increase for all participants.

$$\text{Percentage Increase} = (HF_{relaxation} - HF_{Stress\,Test}) \times 100 \quad (1)$$

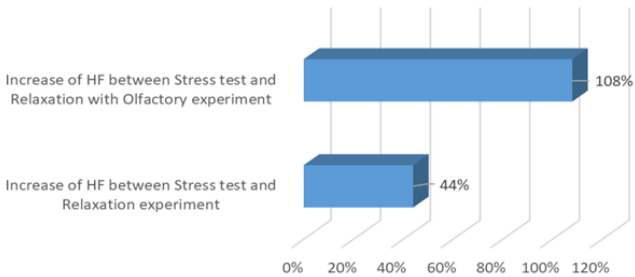

Fig. 3. HF percentage increase between Stress Test and the Relaxation experiment

To evaluate the impact of olfactory stimulation on relaxation, we measured the increase of the High Frequency component under two distinct conditions: relaxation without olfactory stimulation and relaxation with olfactory stimulation.

Relaxation without Olfactory Stimulation: for this experiment, participants were subjected to a relaxation session without any olfactory stimuli following a math stress test. The analysis revealed a 44% increase in the HF component when compared to the HF values recorded during the math test. This significant increase suggests that even without olfactory input, VR-based relaxation techniques can effectively enhance parasympathetic activation, promoting a state of relaxation.

Relaxation with Olfactory Stimulation: On the other hand, when olfactory stimulation was introduced during the relaxation session, the HF component saw an even more substantial increase. As shown in Figure 3, there was a 109% rise in HF values from the math stress test to the relaxation with scent condition. This big increase underscores the potential additive effect of sensory stimulation through pleasant scents in boosting the relaxation response beyond VR-based relaxation techniques alone.

### B. Statistical Comparison and Significance

To determine the statistical significance of these observations, a paired sample t-test was conducted. The paired t-test was chosen because it compares two conditions within the same participants, reducing individual variability and providing a more precise measure of the effect of olfactory stimulation. This approach ensures that any observed differences in HF values are directly attributed to the change in conditions rather than personal differences such as baseline stress levels, age, or health conditions. This way, we isolate the effect of olfactory stimulation from other personal factors. The test compared the mean increases in HF values between the two relaxation conditions. The resulting p-value of 0.002 is well below the standard threshold of 0.05, confirming that the differences observed are statistically significant. Therefore, we can conclude with confidence that relaxation sessions augmented by olfactory stimulation led to greater parasympathetic activation compared to sessions without such stimuli. This finding highlights the potential of integrating olfactory elements into relaxation practices to enhance their effectiveness. A subjective rating scale (e.g likert scale 1 to 10) was used in measuring perceived stress reduction and relaxation in the olfactory and non-olfactory conditions. A paired t-test was used because the study involved the same participants experiencing both conditions (olfactory and non-olfactory), making the data paired and dependent. This test is ideal for within-subject designs as it accounts for individual differences and focuses on the mean difference between conditions. The mean relaxation score is slightly higher for the olfactory condition (7.68) compared to the non-olfactory condition (7.29). This suggests that participants generally perceived the olfactory-enhanced VR environment to be slightly more relaxing than the one without scent. However, the difference is quite small (only 0.39 points on a 1-10 scale). The standard deviation is somewhat higher in the olfactory condition (2.25) compared to the non-olfactory condition (1.94), indicating that there was more variability in how participants rated their relaxation in the olfactory condition. This can also suggest that not everyone experienced the same level of relaxation from the scent addition. In other words, the perceived participants' relaxation experiences with scent varied more than their relaxation experiences without scent.

The increased variability (SD) in the olfactory condition could imply that individual differences (e.g., personal preferences, scent sensitivity) play a role in how effective the olfactory display is for different users. The p-value ($p=0.371$) is greater than $p=0.05$, the difference in relaxation ratings between the olfactory and non-olfactory conditions is not statistically significant. This means that while the olfactory condition had a higher mean relaxation score, the difference is likely due to random variation rather than a true effect of the olfactory display.

*C. Subjective Evaluation Analysis*

The study also investigated participants' subjective opinions on the potential of olfactory-enhanced VR as a relaxation tool. When asked, "Would you use this method as a relaxation technique?" 20 participants (71.4) responded "yes," 4 (14.3%) responded "no," and 4 (14.3%) responded "maybe." These responses suggest a generally positive attitude toward the use of olfactory-enhanced VR for relaxation, with the majority expressing willingness to use it as a relaxation method in the future. Comparing the physiological effects ($p=0.002$) and the perceived effect ($p=0.371$), the physiological data shows a statistically significant effect of the olfactory display. This means that even though participants didn't report feeling significantly more relaxed, their bodies responded differently when the olfactory display was present, showing a measurable reduction in stress or increase in relaxation. The reason for these differences in results could be due to self-reporting limitations (e.g., Likert scales), which can sometimes fail to capture subtle changes in states like stress and relaxation. Participants may not have been fully aware of how relaxed they felt, or they may have been influenced by other factors (e.g., expectations, bias, mood). The multimodal experiment design shows importance since the subjective data alone would not have revealed the significant impact that olfactory stimuli had on participants' relaxation. Future studies could investigate the duration or types of scents that might bridge the gap between perceived and physiological effects. These findings have implications for VR and digital wellness applications, as olfactory stimuli could be incorporated into experiences aimed at reducing stress, even if users don't always perceive the immediate impact.

## VI. CONCLUSION

This study investigated the effects of olfactory displays in VR on stress reduction and relaxation, using self-reported and physiological measures. While scent did not significantly affect perceived relaxation, physiological data showed a notable stress reduction with olfactory stimuli, suggesting subconscious benefits. Most participants favored using this method for relaxation, highlighting the value of multisensory integration in VR for well-being. Future research could explore personalized scents, long-term effects, and additional sensory modalities to enhance VR's effectiveness.